\documentclass[11pt]{article}
\usepackage{amssymb,amsmath,graphicx,float,wrapfig,fancyhdr,lastpage,url,subcaption,textcomp,caption, authblk}
\usepackage[margin=1in]{geometry}
\usepackage[backend=bibtex]{biblatex}
\usepackage{float}
\usepackage{hyperref}
\hypersetup{
    colorlinks = true,
    allcolors = blue
    }
\begin{document}
\title{\textbf{Modeling smart growth of cities through entropy and logistics
}}
\author[1]{James Flamino\thanks{flamij@rpi.edu}}
\author[2]{Alexander Norman\thanks{normaa@rpi.edu}}
\author[2]{Madison Wyatt\thanks{wyattm@rpi.edu}}
\affil[1]{Department of Physics, Applied Physics, and Astrophysics, Rensselaer Polytechnic Institute, 110 8th St, Troy, NY 12180, USA}
\affil[2]{Department of Mathematical Sciences, Rensselaer Polytechnic Institute, 110 8th St, Troy, NY 12180, USA}
\date{June 14, 2017}
\maketitle
\section*{\hfil Abstract\hfil}
\indent
\indent We introduce a predictive algorithm for the smart growth of cities with populations upward of 100,000, allowing for extensive simulations of growth plans and their effects upon an urban populous. A smart growth metric is calculated to evaluate the progress of a city at each phase of its adaptation of the growth plan, which is measured using a weighted entropy method. The predictive algorithm itself is built from a unique differential model, which calculates the growth of a city from smart growth proposals that are individually assessed by a logistic weight model. These proposals are then sorted based on effectiveness and efficiency observed from the simulations, giving insight into the best approach to providing the target cities with a hopeful future. \textit{Original paper written for COMAP's 2017 ICM competition.}
\\
\newpage
\section{Introduction}

\subsection{Context and Motivation}
\indent
\indent The nature of our world is evolving. The industrial revolution of the late 18th century brought upon a great migration into cities, and the trend of urbanization was born. Not only has this trend continued until today, but it continues at an ever increasing rate; it is expected that 66\% of the world will be living in cities by 2050. However, in the wake of this rapid progress, some measures have fallen to the wayside. Some cities either grow too hastily and do not have the economic growth to support it, or have outdated methods for prosperity, all the while continuing to grow in size. Some have not considered the environmental impact and have in turn created habitations that are naturally unsustainable. Some have allowed portions of their community to fall below standards in living and/or in representation. All three of these functions, if not properly maintained, can cause a city to be unequipped for changes that the future brings.\\ 

\indent Smart growth is a movement where cities endorse programs and initiatives that improve three elements: economic prosperity, social equity, and environmental sustainability \cite{SmartGrowth}. In this, there are ten principles given to focus the development of the movement. They serve to further the three E\textquotesingle s with a more detailed and focused set of standards. Cities that adopt smart growth programs are better prepared for the future, despite any uncertainty that could bring. In this paper, we examine the data of cities around the world with midsize metropolitan city populations and analyze the current state of affairs for each by crafting a unique metric rooted in weighing the importance of smart growth in the overall health of a city. Using this metric, two cities with great potential to benefit from the tenets of smart growth are identified: Anchorage, USA and Bissau, Guinea-Bissau. We measure the growth of these cities unattended, then while enacting particular smart growth policies specifically designed to meet the needs of the target cities. The results of the simulations for the policies are measured using the smart growth metric and ranked for comparison. The results allow us to identify the pressing needs of the target cities, and how best to address them with respect to furthering smart growth. \\

\indent In the case of the test cities (Anchorage, USA and Bissau, Guinea-Bissau), the simulation was able to show that enacting particular smart growth policies would grant the cities a noticeable boost in growth, as well as slower degradation of the quality of well-being over the next four decades, especially if the initial enacted policies targeted diverse developmental blueprints, local culture, adaptive urban growth restrictions, and multi-transportational systems, as ranked by the simulation. In addition to these positive results, the simulation itself benefits from having ubiquitous base models, allowing for similar testing to be performed on many other cities for a range of growth plans.

\section{Models}
\indent
\indent Now to build a thoroughly objective, purely math-based simulation to measure the effectiveness of the growth plan policies, specifically, the process was broken down into three models, each rooted in their own fields of mathematics. The then synthesized model serves to take a growth plan initiative, quantify it, and assess its effect on satisfying both the ten principles of smart growth and the three E\textquotesingle s of sustainability. Finally, these effects are integrated to define the resulting growth index of the city due to the implemented initiatives. This multi-step, dynamic system allows a variety of growth plans to be tested to determine the predicted success before commitment and implementation. \\

\subsection{Determining the Sustainability Metrics}
\indent
\indent To consider the details behind the three E\textquotesingle s of sustainability to design a metric, we decided to delve deep into the concepts of being sustainable, and acknowledge that this truly was something of a holistic problem.\\

\indent  Cities naturally involve a myriad of complex interconnected processes, and thus generate a rather large set of data and metrics to evaluate themselves, and none of that data is useless, as interconnected and nuanced as it is. Understanding this, the model first employs the Entropy Weight Method to measure the plethora of information to compute the metric that could be used to measure the sustainability of a city.\\

\indent Generally, entropy is a concept of Thermodynamics, used to measure the disorder of a system. However, the same basic principle exists in information theory, in which the lower the entropy of information for a source, the smaller the utility of said information. This can be applied to a group of indicators, which are then run through this method to determine a value to the system where they lie. Essentially, the entropy method can compute a plethora of indicators involved in a targeted system and evaluate them for their use. \cite{Entropy, Entropy2} This is, of course, a highly useful tool for evaluating the three E's of sustainability, allowing us to consider as much data as we need to inclusively represent each with a unique index.\\

\subsubsection{Evaluation System}
\indent

\indent For a city to satisfy the three E\textquotesingle s of sustainability, it must be Economically Prosperous (EP), Environmentally Sustainable (ES), and Socially Equitable (SE). In order to create a well-rounded metric, a system of indicators that symbolize a component of each of the three sustainability metrics was assembled in order to appropriately represent the significance of each metric after being factored into the entropy method. There are some considerations that truly attribute to each sustainability metric:

\begin{enumerate}
  \item The indicator's change must have a direct effect on the E that they are representing. (i.e. forest density and coverage directly represent a environmentally sustainable system)
  \item The indicator must have complete digital records stretching back to 2000 in order to allow for time-stressing of the resultant metric.
  \item The indicator's type must have data available to not only our two target cities, but also to all similarly sized and located cities.
  \item The indicator must be state- or city-specific in order to assess the most localized data.
\end{enumerate}

\begin{figure}[H]
\centering
  \includegraphics[width=5in]{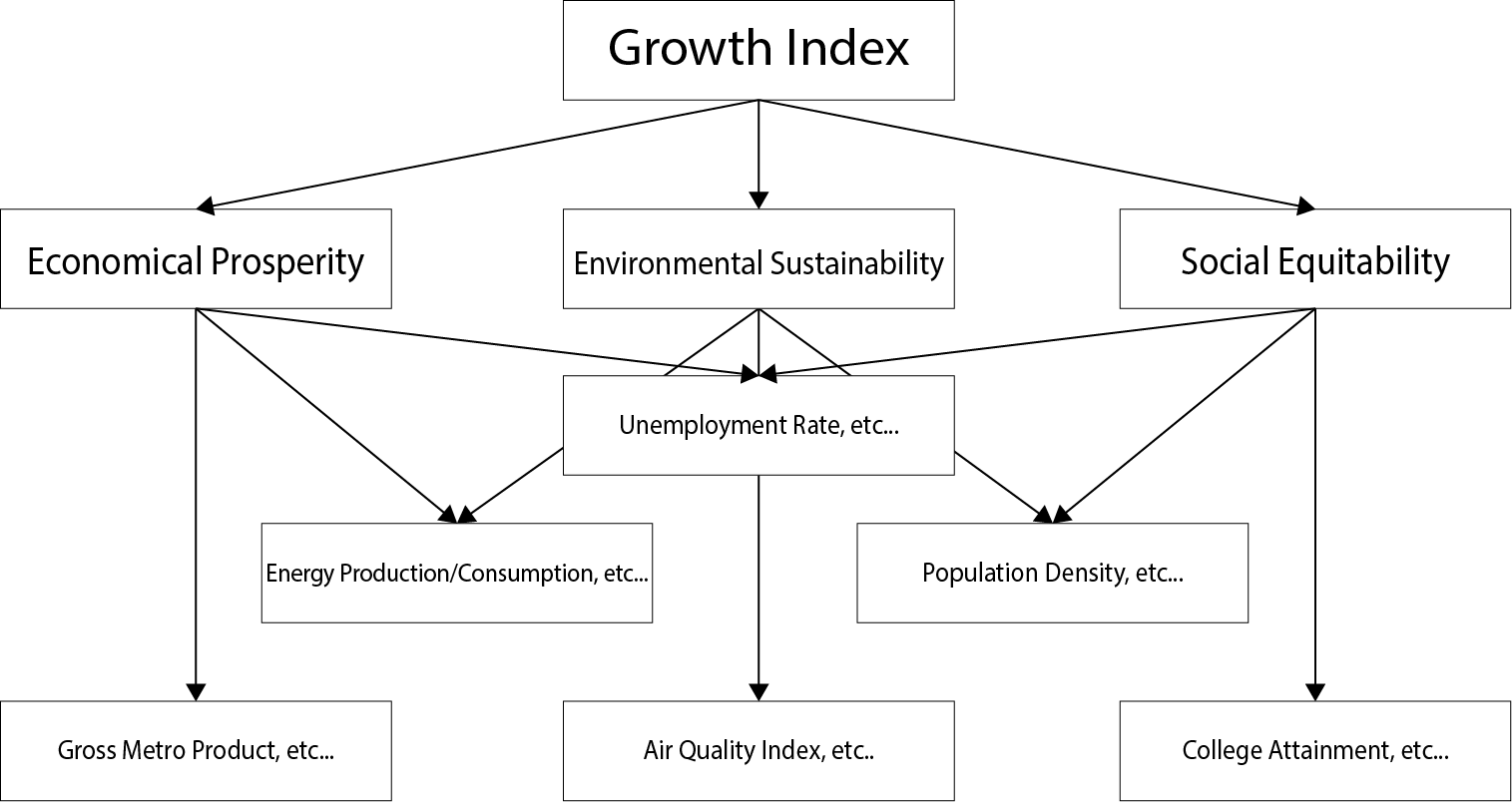}
  	\caption{Network flow model describing the indicator criteria.}
  \label{fig:pbpT0}
\end{figure}

\indent Using these criteria, we were able to collect 150 unique indicators \cite{ClimCalc,EduAt,GDPMet,GBCN,Rank,World}, available openly to cities around the world. These indicators were split up and assigned (and cross-assigned) to one or more of the three categories.\\

\indent This distribution of indicators feeds into the three E's which will each have a special index, all of which feed into one Growth Index, our metric for measuring the overall success of a city. Now, it is important to note that for our model we used more than just two cities - a model that applies to merely two cities is poorly justified on a theoretical level. In order to gain an understanding of overall Smart Growth, we considered a wide sample bin of cities with populations between $100,000$ and $500,000$. In addition to that, we scored all cities together, regardless of continental location in order to gain relative perspective on the progress of each city throughout recent time.\\
\begin{figure}[H]
\centering
  \includegraphics[width=6in]{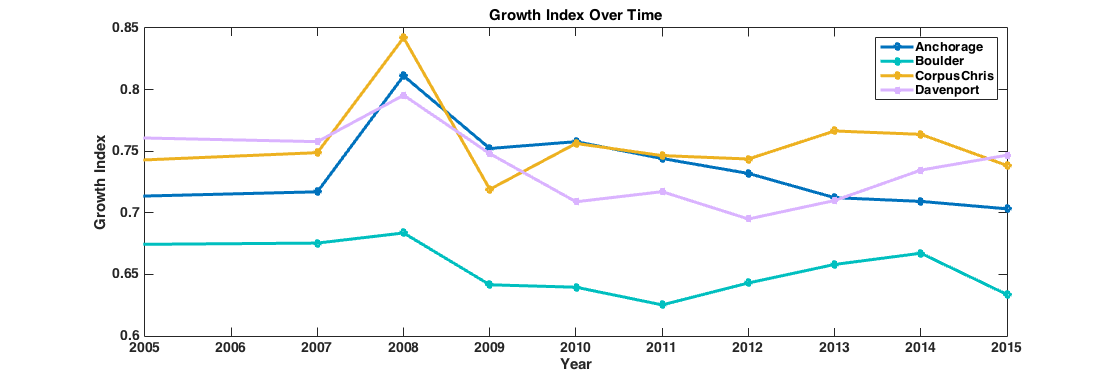}
  \caption{The growth index was calculated for a number of cities with a midsized population to see the current trends, four US cities are shown here. Notice the downward trend of Anchorage, USA in the recent years.}
  \label{fig:pbpT0}
\end{figure}
\subsubsection{Definition}
\indent
\indent Considering the categories outlined in Figure 1, all data was discretized and organized by indicator type and city sample. Thus we can say a single value for an indicator is $x_{ij}$ where $i$ is the indicator for city $j$ where the ranges are $i=(1,2,...,n)$ and $j=(1,2,...,m)$ for all $n$ indicators and $m$ cities.\\

\indent During the initial assessment of indicator values for the city samples, we noted that some indicators were not comparable or scalable to others of the same type when normalization was attempted. In our search for indicators that met the criteria perfectly, we forced an issue inherent of vector normalization: Important variation between indicators both positive and negative is often lost when normalization is attempted, which renders those particular indicators useless. In order to avoid this, the Z-score standardization method was used to normalized all data. \cite{Entropy2} In particular, the Z-score deals well with discrete data, especially for data sets where there is no clear maximum or minimum. And so, for a data value of $x_{ij}$, we say

\begin{equation*}
    r_{ij} = -\frac{x_{ij} - \bar{x_{i}} }{ \delta_{i} }
\end{equation*}

\indent Where $r_{ij}$ is the standardization of $x_{ij}$, and $\bar{x_{j}}$ and $\delta_{i}$ are the mean and the standard deviation of the $i$th indicator of the system, respectively.\\

\indent However, despite the usefulness of Z-score standardizing data instead of just using vector normalization, inaccuracies in the subsequent calculations could arise from the positive and negative distributions of $r_{ij}$. So, all standardized data was transformed into a positive range above 0.

\begin{equation*}
    r'_{ij} = r_{ij} + \phi
\end{equation*}

\indent Where $r'_{ij}$ is the standardized value $r_{ij}$ translated by $\phi$ where $\phi > |min(x_{i})|$. From here the specific value's weight was determined by

\begin{equation*}
    f_{ij} = \frac{r'_{ij}}{\sum_{j=1}^{m} r'_{ij}}
\end{equation*}

Then, the indicator's entropy was calculted by

\begin{equation*}
    H_{i} = -k \sum_{j=1}^{m} f_{ij} \cdot ln(f_{ij})
\end{equation*}

Where $k=\frac{1}{ln(m)}$, which is used to normalize the entropy, $m$ being the number of cities. This is immediately followed by the calculation of the indicator weight, which is

\begin{equation*}
    w_{i} = \frac{1 - H_{i}}{\sum_{i=1}^{n} (1 - H_{i})}
\end{equation*}

Where $n$ is the number of indicators. Finally, this weight was used to calculate an entropy index -- a sustainability metric as we have it -- for each E for the $j$th city sample.

\begin{equation*}
    F_{kj} = 1 - \sum_{i=1}^{n} w_{i}f_{kij} \qquad k = (EP, ES, SE)
\end{equation*}

\indent The three resultant sustainability metrics $F_{kj}$ measure the EP, ES, and SE performance of the $j$th city with respect to all other cities respectively. 

\subsection{Quantifying Initiatives}
\indent
\indent Having gained the tool to evaluate the state of the city through the definition of three sustainability metrics, we then developed a method to calculate the consequences of change through initiatives. In order to effectively assess the quality of a growth plan, we felt it was important to evaluate the plan\textquotesingle s ability to satisfy not only the 3 E\textquotesingle s of sustainability, but also the ten principles of smart growth. This creates relationships between the 3 E\textquotesingle s and the ten principles, as well as relationships between themselves. \\

First, we determined which principles benefited and impaired each of the 3 E\textquotesingle s. A ranking, and weights based on these rankings, were assigned to gauge the varying effects that satisfying a principle would have on each sustainability metric. Conversely, each principle was apportioned by the 3 E\textquotesingle s to establish the weighted intentions of it. That is, the amount a principle is composed of each sustainability metric. Essentially, these two classifications represent column and row weights for the relationship between the 3 E\textquotesingle s and the ten principles.  \\

Additionally, the ten principles are connected. An initiative that changes one principle index may indirectly change another through this change. For this, and founded on the column and row weights determined above, the effects of each principle on the others was concluded and placed in an interconnection weight matrix. These connections are best summarized in the diagram below. \\

\begin{figure}[H]
\centering
\includegraphics[scale=.45]{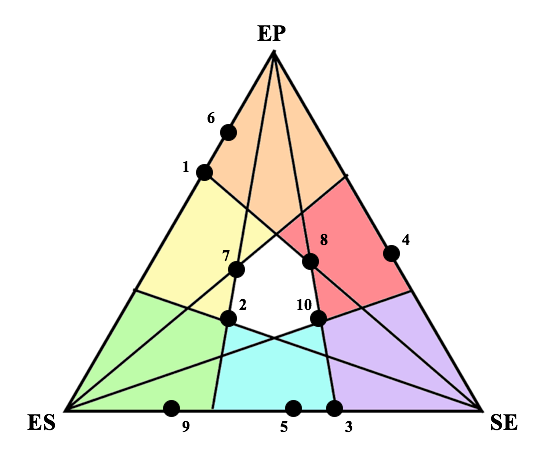} 
	\caption{\small Representation of the relationship between the ten principles and the three sustainability metrics. Each corner represents a metric: Economically Prosperous (EP), Environmentally Sustainable (ES), and Socially Equitable (SE). The area between any two metrics is shown in overlapping color (and shared side), and the area conjoining the three is shown in white. The ten principles were placed as such according to their calculated weights. Relationships between the principles were determined from this as well. 
} 
\end{figure}
\indent Each initiative of a growth plan will satisfy some, but most likely not all, of the ten principles on its own. To measure this, each initiative introduced was decomposed into its contributions to each principle. For example, implementing an initiative to create and maintain a public park in the city would benefit principles 4, 5, and 6 with distributed percentages totaling 100\% for each principle. These classifications were then used to determine the change of each principle index due to the introduction of an initiative. Furthermore, some initiatives may be overall more effective than others in satisfying a principle. For example, setting aside land for agricultural reserve would more adequately satisfy principle 6 than adding natural landscaping to a building front, but both may hold the same distributed initiative percentage for principle 6. Consequently, an effectiveness coefficient, denoted $\beta$, will serve to quantify the quality of the initiative to a principle. \\

\indent The principle indices, the 3 E\textquotesingle s indices, and the overall growth index are defined on scales from 0 to 1, where 0 is not meeting any standards and 1 is satisfying all standards. In calculating the effect of an initiative on a principle, we assume then that the principle index cannot surpass a value of 1. We also assume that as principles are satisfied, additional initiatives have smaller effects. That is, as the value of a principle index approaches 1, each successive change will be smaller.  A general model of this behavior stems from population dynamics.
$$\frac{dx}{dt} = rx \left(1-\frac{x}{K}\right)$$

\indent Where $x$ is the species population at hand, $r$ is the rate of growth, and $K$ is the carrying capacity, the maximum value the population can sustain. Although this model reflects the changing behavior of the principle index, it does not account for the relationship between the ten principles or the effective factor of an initiative. Therefore, a modified logistic model was utilized to characterize the effect of a growth plan initiative on the change in a principle’s index.

$$\frac{dp_k}{dt}=\beta_k i_k p_k \left(1-\sum_{j=1}^{10} w_{kj}p_j \right)$$

\indent Where $p_k$ represents the principle $k$ in question, $\beta_k$ is the effectiveness coefficient of the initiative on the principle, $i_k$ is the initiative contribution to the principle, and $w_{kj}$ is the weighted relationship that a change in principle $j$ would have on principle $k$. Because the model reflects competition between the principles, an influential interaction is denoted as a negative weight and a destructive interaction is denoted positively. It is clear that the quantification of any initiative and the ten principles is consistent and sound. Their transformations are rooted in logic, reasoning, and mathematical consideration. \\

\subsection{Principle Change to Sustainability Metric}
\indent
\indent From here, we must consider how these principles impact our three sustainability metrics, that of economic prosperity, environmental sustainability, and social equitability. To reliably measure this change, we turn to Volterra's Population Equation \cite{NonLinInt}.

\[\frac{du}{dt} = au-bu^2-cu\int^{t}_{0}u(s)ds, \qquad u(0) = u_0\]

Where the constants, \(a\), \(b\), and \(c\) are based on our ten principles. This innovative method evolves the standard population model to better suit the situation at hand. Specifically, the model represents a ``birth rate" term, as to how our principles grow this metric, an ``overcrowding" term, to limit the maximum capacity any city may have, and the final term, described in the literature as a ``toxicity" term, which considers the total of our metric throughout time. This toxicity term slowly tapers off, as the population of our city rises and further resources are required to keep the same levels of our desired values. The equation is then converted to dimensionless form \cite{NonLinInt},

\[\beta \frac{dy}{dt} =  y - y^2-\int^{\tau}_{0}y(z)dz, \qquad y_0 = \alpha\]

by \(\tau =\frac{ct}{b}, y = \frac{bu}{a}, \beta = \frac{c}{a+b}, \alpha = \frac{bu_0}{a}\).\\

\indent However, it becomes important to subtract a time term from the equation because if we take our constants, a, b, and c to be 0, our original equation simply stays constant, which is distinctly unrealistic when considering the infrastructure of a city that decays naturally. Thus we have

\[\beta \frac{dy}{d\tau} =  y - y^2-\int^{\tau}_{0}y(z)dz-d\cdot\tau, \qquad y_0 = \alpha\]

As a technical aside, we can define \(j = \frac{dy}{d\tau}\). This leads to the equation and then definition:

\[j = \frac{1}{\beta}\cdot\left(\int^{\tau}_0j(z)dz-\left(\int^{\tau}_0j(z)dz\right)^2-\int^{\tau}_0\int^{z}_0j(i)didz-d\cdot\tau\right)\]

\[Aj = \frac{1}{\beta}\cdot\left(\int^{\tau}_0j(z)dz-\left(\int^{\tau}_0j(z)dz\right)^2-\int^{\tau}_0\int^{z}_0j(i)didz-d\cdot\tau\right)\]

Together this comes to represent a simple equation,

\[Aj=j\]

As our operator, A is continuous along j, on any fixed interval, we can show, by Brouwer’s Fixed Point Theorem \cite{Func}, that a solution to this integral equation exists if we approximate our function space to be finite dimensional, and thus a solution to our differential equation exists. Further work may be conducted using Scarf’s constructive proof \cite{Brouwer2,Brouwer3} of Brouwer’s Fixed Point Theorem, however at the present time this is beyond the scope of the paper.  The limit of this equation, where \(c=0, d=0\), approximates a function of the form \(f(t) = \frac{e^t}{e^t+C_1}\). Similarly, if we take \(c\) to be suitably large this will approximate a \(sinh\) function. However, no analytic solution is easily attainable, so a fourth order Runge Kutta method was used to compute how each index changes, treating the one equation as a system of two like so.
 
\[\frac{df}{dt} = y\]

\[\frac{dy}{dt} = y-y^2-f-t\]

Solving these two sets of equations in parallel over five thousand points in our interval for each simulation returns a reasonable approximation of both \(f\), and, more importantly, \(y\). Which, of course, given the correct parameters, can adequately represent each of our three E's and how they change over time.\\

\indent An important thing to discuss here is how precisely our constants in this model are formulated. For simplicities sake, and as a good first order approximation, we take each to be a linear combination of our 10 principles, discounting any cross correlation between them. Then, we evaluate how much impact, positive and negative, and taking into account population growth, these principles have on our three metrics, we normalize this to one when you add the scaling across the metrics for each principle. Then we consider how impactful each principle is \textit{ with respect to each individual metric}, normalizing the sum of all ten principles to 1 across each metric. These two numbers are then multiplied, giving an overall weight.

\subsection{Calculation of Growth Index}
\indent
\indent Finally, with a model that can comprehensively represent each sustainability metric over time, it makes sense to attempt to quantify those sustainability metrics as a total, as it will illustrate the overall success of a city over a period of time more concisely.\\

\indent And so we introduce the Growth Index (GI), our central metric for measuring the overall smart growth of a city. This is defined as

\begin{equation*}
GI = W \cdot F = 
    \begin{bmatrix}
        \gamma_{1} \\
        \gamma_{2} \\
        \gamma_{3}
    \end{bmatrix}^T
    \cdot
    \begin{bmatrix}
        F_{1} \\
        F_{2} \\
        F_{3}
    \end{bmatrix}
\end{equation*}

Where $\gamma_{1}$, $\gamma_{2}$, and $\gamma_{3}$ are distributive weights and $F_1$, $F_2$, and $F_3$ are the growth indices for the 3 E\textquotesingle s, respectively. These are very important, as they are weighted to equally represent each $F_{k}$ entropy index fairly, preventing a city that does not excel in all three E's from standing out. These weights are calculated through the following equations

\begin{equation*}
    q_{i} = \frac{F_{i} - \frac{M}{2}}{|F_{i} - \frac{M}{2}|}
\end{equation*}

\begin{equation*}
    F'_{i} = (\frac{M}{2} - F_{i}) \cdot \left(\frac{1}{2}\right)^{q_{i}} + \frac{M}{2}
\end{equation*}

\begin{equation*}
    \gamma_{i} = \frac{F'_{i}}{\sum_{i=1}^{3} F'_{i}}
\end{equation*}

Where $M = max(\{F_{1}, F_{2}, F_{3}\})$, $F'_i$ is the distributive value of the index, and $i$ is the $i$th $\gamma$, where $i = (1,2,3)$. These calculations in total result in the growth index of a city. It is used to quantify the health of the city in terms of sustainability.

\section{Analysis of Cities}
\subsection{Current State of the Cities}
\indent
\indent The two cities that are detailed and analyzed throughout the remainder of this paper are Anchorage, USA and Bissau, Guinea-Bissau. With the sustainable indices and an overall comprehensive metric for smart growth in a metropolitan area on hand, we embarked on the task to analyze as many cities as we could in order to gain a detailed understanding of the progress of cities from the USA to Africa. As data allowed, we were able to analyze 38 different US cities and 8 different African cities, and we were able to pinpoint two particular cities of interest: Anchorage, USA, and Bissau, Guinea-Bissau. Despite a difference in base standards, both of these cities face a similar problem: At their current rate, their well-being as indicated by the Growth Index is falling, and they will face many future problems if not corrected. In addition, we found that both of these cities in fact beneﬁt from smart growth initiatives.\\

\subsubsection{Anchorage, USA}
\indent Anchorage, USA has experienced a decrease in its growth index over the past decade. It has seen a decrease in economic fortune in the past few years, even after many years of moderate success \cite{Gover}. It’s homelessness problem has reached a new high, creating not only a divide in the population, but an environment ill-conducive to development \cite{Smartgrowth2}. It has boasted an air quality index dangerously higher than the national average, and well above a standard of clean air, and continues to lead in US crude oil production \cite{ClimCalc,Anchorage} . However, in the past 6 months, it has begun to plan for the future, attempting to adopt a smart growth plan for itself \cite{Smartgrowth2}. In this time, it has published an incentive handbook, required progress reports from its development authority, worked to redefine development, added more foot patrol downtown to increase safety, begun a community homeless program, and worked to designate a trail and rain garden. We examined, first, the state of the city before these initiatives. Then, using our model and metric, we assess how well these initiatives will work as time progresses.\\

It is clear by the current downward trend and the current state of Anchorage\textquotesingle s growth index, that without intervention, the state of the city will fall considerably. The city suffers most in social equity, then economic prosperity, and finally in environmental sustainability. The initiatives currently presented were placed in the model and the results over time were seen in the figure below. Individually, each initiative either causes the growth index to decay immediately, or rise and fall quickly before leveling off around the initial point or below it. If all the initiatives are enacted, the system rises to a growth index of below 0.9 before falling back down quickly. Left alone, these initiatives would not allow the city to prosper at a sustainable level or account for the population increase it is expecting in the long run. This is seen in the figure. \\
\begin{figure}[H]
\centering
\includegraphics[scale=.4]{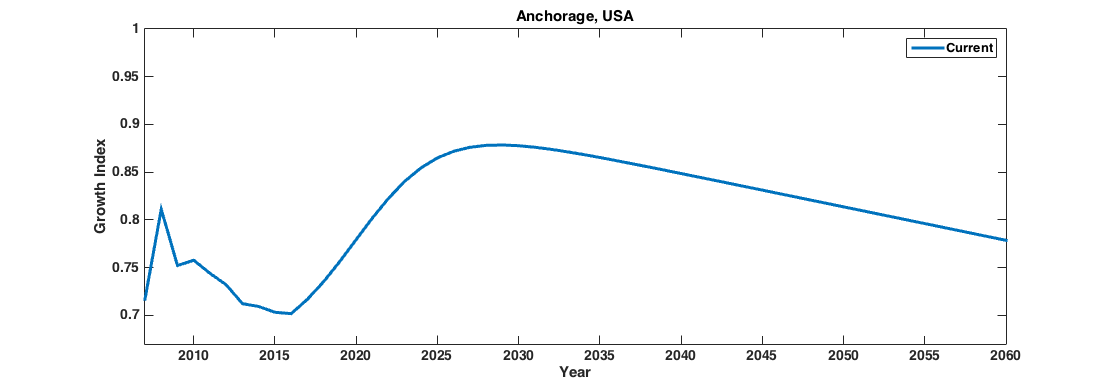} 
\caption{\small Past growth indices and future predictions based on the current state of Anchorage, USA.} 
\end{figure}

\subsubsection{Bissau, Guinea-Bissau}
\indent
\indent Bissau, Guinea-Bissau is beginning to see a rise in its economic future. Its Gross Domestic Product (GDP) has grown consistently by 5\% in the last two years since the election of a new president after the country was plagued with civil war beginning in 2012. There is a large central market for goods exchange in the downtown area, as well as beautiful coastlines for economic and environmental growth. However, the country has not had a long period without political unrest since its independence in 1974. The economy is mostly dependent on cashew and rice production outside of the city and very few tourists visit Bissau \cite{Livelihood}. There is rising inequity with the poverty rate increasing to unprecedented levels. Much of the city is divided between two standards of living, one of which is well below any standard at all \cite{GBCN}. The city has a few plans in place to begin alleviating these troubles. Currently, they discuss setting aside a national park, demobilizing portions of the soldiers, and economically empowering poor farmers, especially women \cite{Livelihood}. We again analyzed the state of the city before these plans and then ran the initiatives through our model to assess their effectiveness.\\

\indent The current state of the city of Bissau is projecting increase in social equity, decrease in environmental sustainability, and slight, if any, increase in economic prosperity. Although the implementation of the initiatives did serve to further the metrics for a few years and ultimately increase the growth index of the city, the index is still not at a point that is sustainable for future growth. Individually, each initiative was not enough to cause the city to improve each metric wholly. All together, the increase was quick and unsustainable for growth in the city. This is seen in the figure below.\\
\begin{figure}[H]
\centering
\includegraphics[scale=.4]{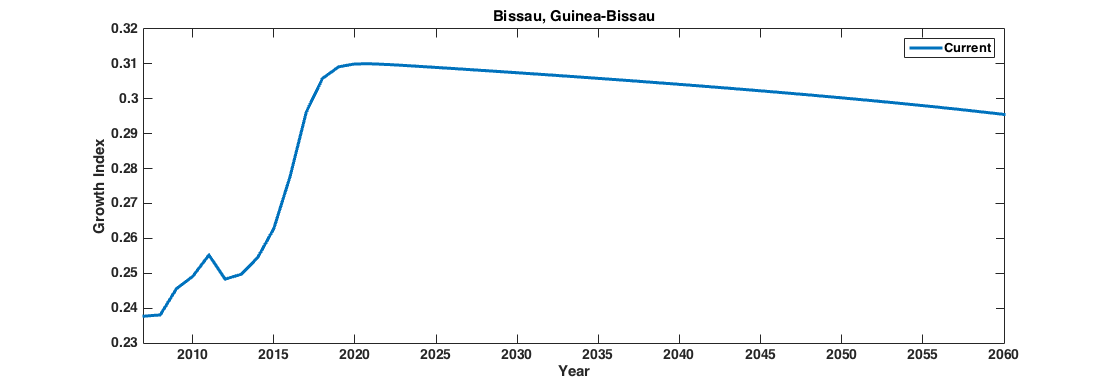} 
\caption{\small Past growth indices and future predictions based on the current state of Bissau, Guinea-Bissau} 
\end{figure}

\section{Initiative Plan: Change For The Better}
\indent
\indent The concept of smart growth is founded on ten principles of progress and three tenets of sustainability. These core beliefs form plans and programs to create better futures for our cities and communities. It is from these values and our comprehensive analysis of the current state of growth plans for Anchorage, USA and Bissau, Guinea-Bissau that we form an inclusive plan, with an intention for focused and detailed variations for each city.\\ 

\indent Having analyzed both the state of each city as it pertains to economic, social, and environmental growth, as well as the current growth plan to improve those and its consequences, we have introduced a more suitable plan that follows with ten Smart Growth Initiatives (SGIs).
\begin{enumerate}
	\item Focus on small areas for improvement
	\begin{itemize}
		\item Improve long-term sustainability by utilizing short-term, but steady, independent programs
\end{itemize}
\item Incorporate local culture
\begin{itemize}
\item Art, sculptures, history, important aspects
\item Deepen community sense by recognizing the people present 
\end{itemize} 
\item Multi-transportational system
\begin{itemize}
\item Public transportation, bike access
\item Walkways, bike lanesequal opportunity
\end{itemize}
\item Consider inherent hindrances and adjust accordingly
\begin{itemize}
\item Take into account geographical features, political situations, and other uncontrollable factors
\end{itemize}
\item Work to Reduce, Reuse, Recycle instead of Rebuild or Restart
\begin{itemize}
\item Apply environmentally sustainable practices 
\item Assess the state of the city road and building design before enacting plans
\end{itemize}
\item Pursue outside financing and grants
\begin{itemize}
\item Encourage developmental leadership to use community and regional resources to fund improvement projects
\item Create a clear, concise objective for proposals
\end{itemize}
\item Exploit markets and benefits of the city
\begin{itemize}
\item Strengthen well-being 
\item Examine pre-existing conditions for strong components
\end{itemize}
\item Monitor restrictions and work to reflect the goal of the growth 
\begin{itemize} 
\item Adapt to current conditions
\end{itemize}
\item Implement unique and diverse developmental blueprints
\begin{itemize}
\item Tailor design proposals to the city at hand
\item Outline layouts that reflect the community
\end{itemize}
\item Introduce or enhance inclusion programs
\begin{itemize}
\item Strive for fair treatment and services for the community
\item Create or expand on existing programs designed for justness
\end{itemize}
\end{enumerate}
\indent
\indent This growth plan is tailored to the specific needs of the two communities in Anchorage, Alaska and Bissau, Guinea-Bissau. Both of these cities are currently struggling with social equity. Anchorage has an immense homelessness problem, low racial diversity, and few initiatives in place to combat these issues \cite{Anchorage}. Bissau is, too, struggling with 49\% of the populating below the poverty line \cite{Livelihood}. Economically and environmentally, the two cities are similarly situated, relative to itself, in that their focuses for growth are both going to be on social, then economic, then environmental.\\

\indent It would be advantageous for both cites to enhance what already works best for each. Both cities are expecting growth in the upcoming future -- Anchorage in population and tourist attraction, and Bissau in Gross Domestic Product (GDP), a measure of their economic health \cite{GBCN,Anchorage}. This plan also accommodates the geography, political condition, and funding availability of each city. Bissau more than Anchorage, but each city will need to seek outside sources to fund this growth project. Finally, this plan reflects a feasible solution for each city in its own manner. Our model accounts for prioritization in the sustainability metrics for each city, so the amount of resources expelled for each of these initiatives is customized to the community and current situation of each city.\\ 

\subsection{Initiatives in Model}
\indent 
\indent These ten initiatives were then run through all components of the model to determine how the growth index would change as a result. Each SGI was weighted appropriately and the results are shown best in comparing the future of the city\textquotesingle s growth index with and without these measures taken. 

\begin{figure}[H]
 \centering
  \subcaptionbox{\label{fig4:a}}{\includegraphics[width=6in]{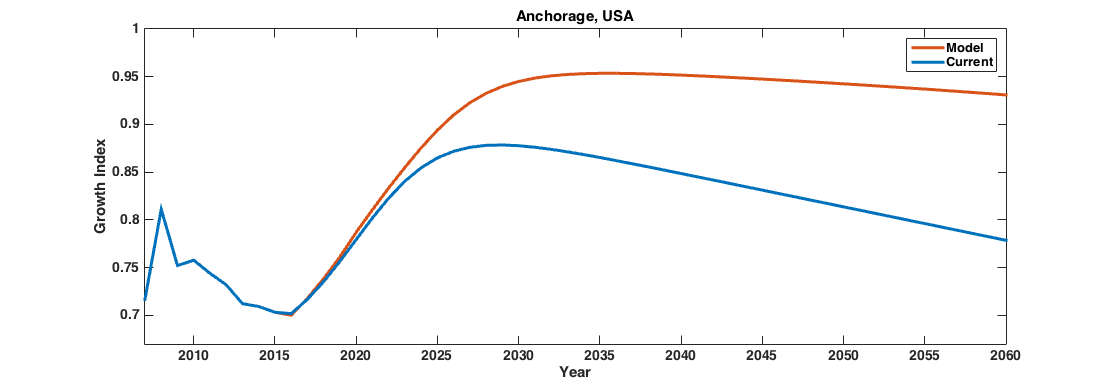}}\hspace{1em}%
  \subcaptionbox{\label{fig4:b}}{\includegraphics[width=6in]{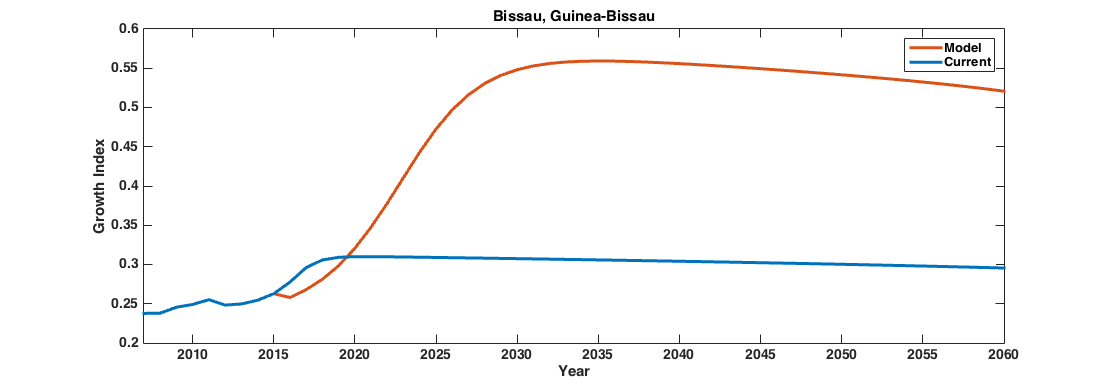}}\hspace{1em}%
  \caption{Past growth index and future predictions for (a) Anchorage and (b) Bissau. The model curve represents our iniatives and the current curve represents the state of the city as it stands.}
\end{figure}

\indent As seen, both cities see a great increase in their growth indices throughout time. The initial and archival data from real data is included for a ten-year reference into the past. Then, the growth index for the next 40 years is shown for the city as it currently stands, and with our model generated initiative plan. Anchorage, USA reaches a near perfect growth index around 2035 and settles around 0.9 for the duration of the represented data. This is about a 0.15 increase from the index without the plan, as well as showing more promising future prospects because the decay of the index is not as sharp. \\

\indent Moreover, Bissau, Guinea-Bissau sees a dramatic difference in the growth indices. Without the plan, the index settles around 0.3 which is not sustainable for future growth. With the plan, it settles closer to 0.5 with room to improve and/or introduce further initiatives. Note that Bissau\textquotesingle s current plan exceeds our plan for a few initial years. This is due to the nature of their current plans. Bissau has only a few, but large, projects on queue. These projects have a quick effect, but little lasting ability. Thus, our plan overtakes the current initiatives quite quickly. Essentially, it is evident that the Smart Growth Initiatives outlined in our plan are sound, dynamic, and complete. \\

\subsection{City Comparison}
\indent
\indent The ability to enact any number, especially all, of the initiatives outlined in our smart growth plan depends on the city’s ability to fund and support such a project. There may be times in which the city needs to prioritize the programs based on their desired end result, or a city may need to adopt this strategy from the beginning. For example, Bissau may not have the resources to support a full smart growth plan at the moment, and may instead wish to target markedly trouble areas. For this, we have ranked the initiatives. These rankings were determined by running each initiative through the model individually and analyzing the effect on each sustainability index. The results of the analysis are summarized in the colorbar charts. The SGI’s are ranked top to bottom for each metric, so depending on the goal of the city at the time, appropriate initiatives can be selected. \\

\begin{figure}[H]
 \centering
  \subcaptionbox{\label{fig4:a}}{\includegraphics[width=1.02in]{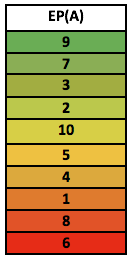}}\hspace{1em}%
  \subcaptionbox{\label{fig4:b}}{\includegraphics[width=1.05in]{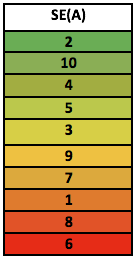}}\hspace{1em}%
  \subcaptionbox{\label{fig4:a}}{\includegraphics[width=1in]{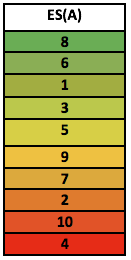}}\hspace{1em}%
\end{figure}
\begin{figure}[H]
 \centering
  \subcaptionbox{\label{fig4:a}}{\includegraphics[width=1.02in]{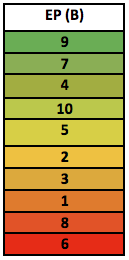}}\hspace{1em}%
  \subcaptionbox{\label{fig4:b}}{\includegraphics[width=1.02in]{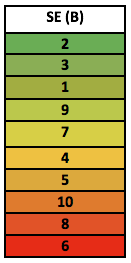}}\hspace{1em}%
  \subcaptionbox{\label{fig4:a}}{\includegraphics[width=1in]{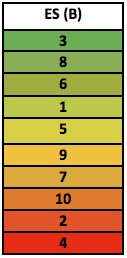}}\hspace{1em}%
  \caption{Initiative rankings for Anchorage (A) (above) and Bissau (B) (below) for each sustainability metric. The initiative with the most potential is at the top of each colorbar, while the least potential is at the bottom.}
\end{figure}

\indent For Anchorage, they may choose to improve environmentally and socially, so it would be more advantageous for the city to begin with SGIs 8 and 2. In comparison, Bissau may choose to focus on economic growth and social equity, so they should begin with SGIs 9 and 2. It is interesting to observe the similarities between the rankings. This notes the all-inclusive nature of the model and its ability to positively affect any city, despite initial conditions.\\

\section{Sensitivity Analysis}
\subsection{Unexpected Perturbations}
\indent
\indent Now that the model has been fully developed, and the simulations show a positive change in the Growth Index for both cities, the model was stress-tested in order to scope out its robustness and effectiveness. The first test of sensitivity analysis considered was to include randomly placed negative perturbations to the Growth Index. These kinds of perturbations simulated the uncertainty of the real world that could noticeably effect the economical prosperity, environmental sustainability, and the social equitability of a city in a detrimental way. For example: a natural disaster could damage the environmental sustainability and economical prosperity of a city, while a political upheaval could negatively effect the economical prosperity and the social equitability.\\

\indent In order to simulate the disasters and their detrimental affects upon a city economically, environmentally, and socially, a $3 \times 3$ perturbative matrix was introduced. This matrix ties directly into the distributive weights $\gamma_{i}$, disturbing their values and affecting the growth indices, which, ultimately, affect the Growth Index itself. The perturbative matrix is of the form

\begin{equation*}
D = 
    \begin{bmatrix}
        \beta_{11} & \beta_{12} & \beta_{13} \\
        \beta_{21} & \beta_{22} & \beta_{23} \\
        \beta_{31} & \beta_{32} & \beta_{33}
    \end{bmatrix}
\end{equation*}

\indent Here $\beta_{ij}$ is a perturbation value between $-1$ and $0$ where the more negative the value, the greater the disturbance. In this form, the matrix can be tuned to affect different parts of the Growth Index in varying degrees. For example, to have a disaster that affects the economy and the environment exclusively, values of $\beta_{ij}$ are generated for $i={1,2}$ with all other values set to $0$. Considering the perturbative matrix in terms of the Growth Index:

\begin{equation*}
W' = D \times W =
    \begin{bmatrix}
        \beta_{11} & \beta_{12} & \beta_{13} \\
        \beta_{21} & \beta_{22} & \beta_{23} \\
        \beta_{31} & \beta_{32} & \beta_{33}
    \end{bmatrix}
    \times
    \begin{bmatrix}
        \gamma_{1} \\
        \gamma_{2} \\
        \gamma_{3}
    \end{bmatrix}
\end{equation*}

\begin{equation*}
    GI' = (W')^{T} \cdot F
\end{equation*}

\indent Where $W'$ is our perturbed weights, and $GI'$ is our perturbed Growth Index.\\

\indent The values $\beta_{ij}$ were not determined entirely through stochastic methods, however. Instead, the magnitudes of the values were weighted differently depending the types of disasters that could potentially happen in the target region, with historical accuracy. To elaborate, consider our initial stress-testing of our perturbation algorithm. For this run, we decided to focus on Bissau, as it\textquotesingle s location in a third-world country, along with it\textquotesingle s historical record, leave it susceptible to potential political unrest, not to mention the occasional flood or locust infestation. Studying the past 50 years of history for Guinea-Bissau \cite{GBCN,Disas} carefully, we were able to set up a Poisson distribution for political disasters, natural disasters, and economical disasters for three degrees of severity for the next 50 years. So, these distributions worked to indicate the probability of a disaster triggering at each time step in our differential model. Once a disaster of a particular severity was triggered, the appropriate row (or rows) of values in the perturbative matrix were randomly determined, weighted toward either $-1$ or $0$ depending on the severity of the disaster as indicated by the distribution. With the perturbative matrices generated for the appropriate time steps, the results were fed into the Growth Index.\\

\begin{figure}[H]
\centering
\includegraphics[scale=.4]{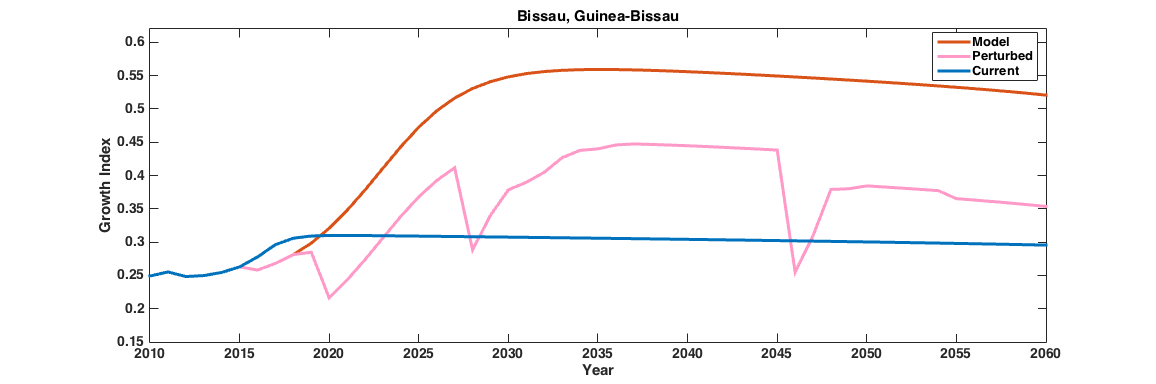} 
\caption{\small Growth index over time for Bissau. The current curve represents the index of the city as it stands, the model curve represents the index over time with our growth plan, and the perturbed curve represents the growth index given unexpected, random disasters.} 
\end{figure}

\indent The results of our initiative through time with unexpected events is seen in the figure. In the perturbed curve, Bissau undergoes major disasters in the years 2020, 2028, 2046. These events set back the Growth Index greatly, but regardless of the severity of each disaster, our initiatives continue to have a positive effect on Bissau, allowing for recovery in the Growth Index in a short period of time (albeit somewhat less in magnitude than previously projected). This result goes to show that the model along with the choice selection of initiatives retain the capacity to grow a city\textquotesingle s smart growth potential regardless of possible disasters that may arise in the future. Our growth plan, in short, is robust and effective in the face of the real world.\\

\subsection{Dramatic Population Increase}
\indent
\indent The model is also very robust when taking population change into account. For example, consider that each city will increase its population by 50\% by 2050. The differential model accounts for population change over time, in part represented by our ``toxicity" term. This term quantifies adverse effects that scale with population, including societal unrest, discohesion due to policies, or increasing maintenance, etc. Accounting for a specific population growth can be, and is, subsumed into our variables.\\

\begin{figure}[H]
\centering
\includegraphics[scale=.4]{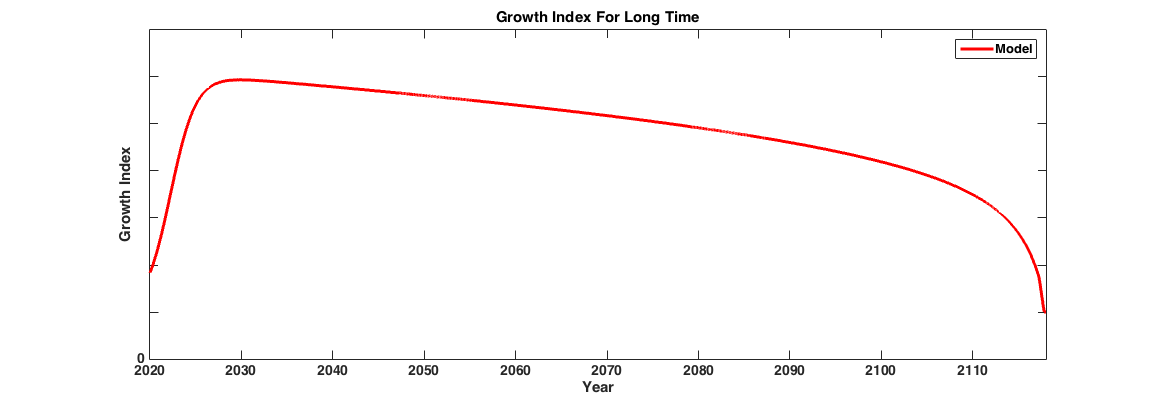} 
\caption{\small As seen, the model accounts for the possibility, and really, inevitability, of population increase by introducing decay over long time.} 
\end{figure}

\indent Indeed, the results produced strongly mirror this fact, with greater decay occurring over longer time scales. As we further the time, the decay becomes quite dramatic, more so than could be accounted for in either the overcrowding term or the decay term. Therefore, the model adequately considers population and not just a static city, which would be physically unrealistic.\\

\section{Conclusion}
\indent
\indent The growth of cities is inevitable. The world population is exploding and urbanization is an accommodating solution. With this expansion comes consequences that affect the environment, the economy, and the social balance. However, the consequences of this progression are avoidable with the implementation of smart programs. Specifically, our ten Smart Growth Initiatives consistently increase a city\textquotesingle s growth index, which takes into account all three sustainability metrics and all ten principles of smart growth. Our model proves that the SGIs are robust, enduring unexpected changes like political, economic, or natural disasters as well as functioning in the face of population increase. It is clear through the extensive data analysis and research that this model is accurate and sound, attempting to remove subjective decisions at all possible points. The basis of the central metric, the overall Growth Index, properly reflects the state of the city at all points, too. It is grounded in real world data and constructed with rigorous calculations. The two cities, Anchorage, USA and Bissau, Guinea-Bissau were visibly shown to improve their growth index with the application of our smart growth plan. Their indices increased considerably and show slower decay as time moves forward. Finally, in the face of perturbations, both indices still showed improvement above the current state of the city.\\ 

\subsection{Strengths and Weaknesses}
\indent
\indent In our model, some strengths and weaknesses come to mind immediately. For the one, the model is completely deterministic and thorough, given starting values it can explain exactly how the three E\textquotesingle s, our metrics, unfold and evolve. There is a consistency throughout the model that becomes important when considering different cities or initiatives. The weights chosen reflect the model, not the input data. It is quite robust, even given the odd source for this model and the rather unorthodox adaptation. In addition our model considers large amounts of real-world data, allowing for it to act upon those values as realistically as possible, with little room for assumptions. Also, it's a rather quick computational model, despite the complexity, taking under a second for each set of parameters. And finally, we can know that a precise solution exists, and unlike in previous population models, we can confidently claim that the solution is non trivial, due to the decay term.\\

On the other hand, our model is explicitly computational and numerical in nature, rather than analytic, and thus is going to have some level of imprecission built in. Similarly, the model is rather undetermined in a and b, with our values of parameters based on reflective reasoning rather than rooted in direct data, simply due to the fact that there are 10 principles and fitting all of them to data to obtain the proper relation would be rather unrealistic. The initiative percentages are self-chosen which is unavoidable because they need to be quantified in some initial way. Finally, as the model acts as a system of differential equations, inputs for both the population function and the antiderivative must be specified for time \(t=0\), while only one input is given. On the relevant time scales, multiple values were tested and it ultimately made little difference, but it's a facet that can certainly be improved.\\

\newpage

\printbibliography

\end{document}